\documentclass[letterpaper, 11pt]{article}
\usepackage[utf8]{inputenc}
\usepackage[margin = 1in]{geometry}
\usepackage[colorlinks, linkcolor=orange, anchorcolor=blue, citecolor=blue]{hyperref}
\usepackage{graphicx} %
\usepackage{xcolor}
\usepackage{siunitx}
\usepackage{indentfirst}
\usepackage[numbers,sort]{natbib}
\usepackage{enumitem}
\usepackage{multirow}

\usepackage{listings}
\usepackage[noframe]{showframe}
\usepackage{framed}
\usepackage{lipsum}
\usepackage{amsmath}
\usepackage{cleveref}

\usepackage[T1]{fontenc}    %
\usepackage{url}            %
\usepackage{booktabs}       %
\usepackage{amsfonts}       %
\usepackage{pifont}         %
\usepackage[most]{tcolorbox}
\usepackage{caption}
\usepackage{xurl}
\usepackage{colortbl}

 {\endMakeFramed}
\definecolor{shadecolor}{gray}{0.75}
\definecolor{lightblue}{HTML}{cfe3f4}

\newcommand{\method}{\text{\textsc{RetroDFM-R}}} %

\title{\method{}: Reasoning-Driven Retrosynthesis Prediction with Large Language Models via Reinforcement Learning}
\date{}

\author{
Situo Zhang$^{1\dagger}$ \quad Hanqi Li$^{1,3\dagger}$ \quad Lu Chen$^{1,2,3,4\ast}$ \quad Zihan Zhao$^{1}$ \quad Xuanze Lin$^5$ \\
Zichen Zhu$^{1}$ \quad Danyu Luo$^{1}$ \quad Bo Chen$^3$ \quad Xin Chen$^3$ \quad Kai Yu$^{1,3,4}$ \\ \\
$^1$ X-LANCE Lab, School of Computer Science, \\Shanghai Jiao Tong University, Shanghai, China\\ 
$^2$Shanghai Innovation Institute, Shanghai, China\\
$^3$ Suzhou Laboratory, Suzhou, China\\ 
$^4$ Jiangsu Key Lab of Language Computing, Suzhou, China\\
$^5$ School of Chemistry and Chemical Engineering, \\Shanghai Jiao Tong University, Shanghai, China\\
$^\ast$ Corresponding author: \texttt{chenlusz@sjtu.edu.cn} \\
}

\begin{document}
\maketitle
\def\thefootnote{$\dagger$}\footnotetext{These authors contributed equally to this work.}

\begin{abstract}

Retrosynthetic planning is a cornerstone of organic synthesis and drug discovery. Yet existing AI methods often rely on pattern matching rather than transferable chemical reasoning, limiting both generalizability and interpretability. Here we introduce \method{}, a reasoning-driven large language model (LLM) for chemical retrosynthesis. Leveraging large-scale reinforcement learning, \method{} moves beyond black-box prediction by coupling improved accuracy with transparent, step-by-step rationale. On the USPTO-50K benchmark, \method{} achieves 60.4\% accuracy without augmentation and 66.1\% with the full inference setup, outperforming previous state-of-the-art baselines. Beyond standard metrics, double-blind expert evaluation further supports the chemical plausibility and practical utility of its proposed pathways. We also demonstrate that \method{} can reconstruct complex, multistep synthetic routes for real-world pharmaceuticals and self-assembled monolayer materials. By making its reasoning explicit and human-interpretable, \method{} addresses a key barrier to trust and supports practical deployment in automated retrosynthetic planning.

\end{abstract}

\section{Introduction}
Retrosynthetic analysis, formalized by Corey~\citep{corey1991logic}, remains foundational to modern molecular synthesis. By iteratively disconnecting a target molecule into readily accessible precursors, it provides a systematic framework for designing efficient synthetic routes. As a result, retrosynthetic analysis has become indispensable in drug discovery and materials science, where rapid access to complex natural products and novel compounds accelerates innovation. Motivated by this central role, computer-assisted retrosynthesis has been pursued for decades~\citep{corey1985computer, todd2005computer, coley2017retrosim}. Early systems largely depended on hand-crafted reaction rules and expert-curated knowledge bases, which are costly to maintain and difficult to scale, and often fail to generalize across the breadth of chemical reactivity.

Recent advances in artificial intelligence have substantially improved single-step retrosynthesis, primarily through two paradigms: graph-based and sequence-based models. Graph-based approaches, including GLN~\citep{dai2019gln}, MEGAN~\citep{sacha2021megan}, and G$^2$Retro~\citep{chen2023g}, typically operate by identifying reaction centers or predicting graph edits to generate precursors. While these methods preserve molecular topology and align with chemical intuition, they often face challenges in computational scalability and continue to rely on restrictive templates or heuristics~\citep{jin2022graph}. In parallel, sequence-based methods frame retrosynthesis as a translation task over linearized molecular strings such as Simplified Molecular Input Line Entry System~(SMILES). Since early applications of recurrent models~\citep{liu2017retrosynthetic}, Transformer-based architectures~\citep{karpov2019transformer, tetko2020state} and optimized representations, including R-SMILES~\citep{zhong2022rsmiles} and EditRetro~\citep{han2024editretro}, have substantially improved predictive performance. However, a critical limitation persists across both paradigms: state-of-the-art models largely operate as end-to-end black boxes, offering little insight into the chemical rationale underlying a predicted disconnection. This lack of explicit reasoning limits trust and constrains the practical utility of AI-driven retrosynthesis in experimental settings.

Large language models (LLMs) have recently achieved strong performance across diverse domains~\citep{brown2020language, achiam2023gpt, jaech2024openai, chen2025dfm}, largely due to their scale and broad pretraining. Progress in Chain-of-Thought (CoT) reasoning~\citep{jaech2024openai, guo2025deepseekr1} further suggests that LLMs can solve complex, multi-step problems with explicit intermediate reasoning, including tasks such as code generation and mathematical problem solving~\citep{DeepMind2025GeminiIMO}. These developments motivate the use of LLMs for retrosynthesis, where the ability to combine broad knowledge with explicit reasoning could improve both generalization and interpretability. However, general-purpose LLMs often struggle with chemical formalisms (e.g., SMILES) and reliable chemical reasoning. Recent chemical LLMs~\citep{zhao2025developing, zhao2024chemdfmx, zhang2024chemllm, yu2024llasmol, tan2026chemmllm} improve domain knowledge via large-scale pretraining on chemical texts, including ChemDFM~\citep{zhao2025developing} and ChemLLM~\citep{zhang2024chemllm}. Nonetheless, these models remain substantially less competitive than smaller, task-specialized approaches on retrosynthesis benchmarks; for example, ChemDFM reports 17.9\% accuracy on USPTO-50K. Moreover, existing chemical LLMs typically treat retrosynthesis as direct end-to-end prediction, lacking both human-readable explanations and the ability to provide practical chemical insight.

\begin{figure}[]
    \centering
    \hspace*{-0.\linewidth}
    \includegraphics[width=\linewidth]{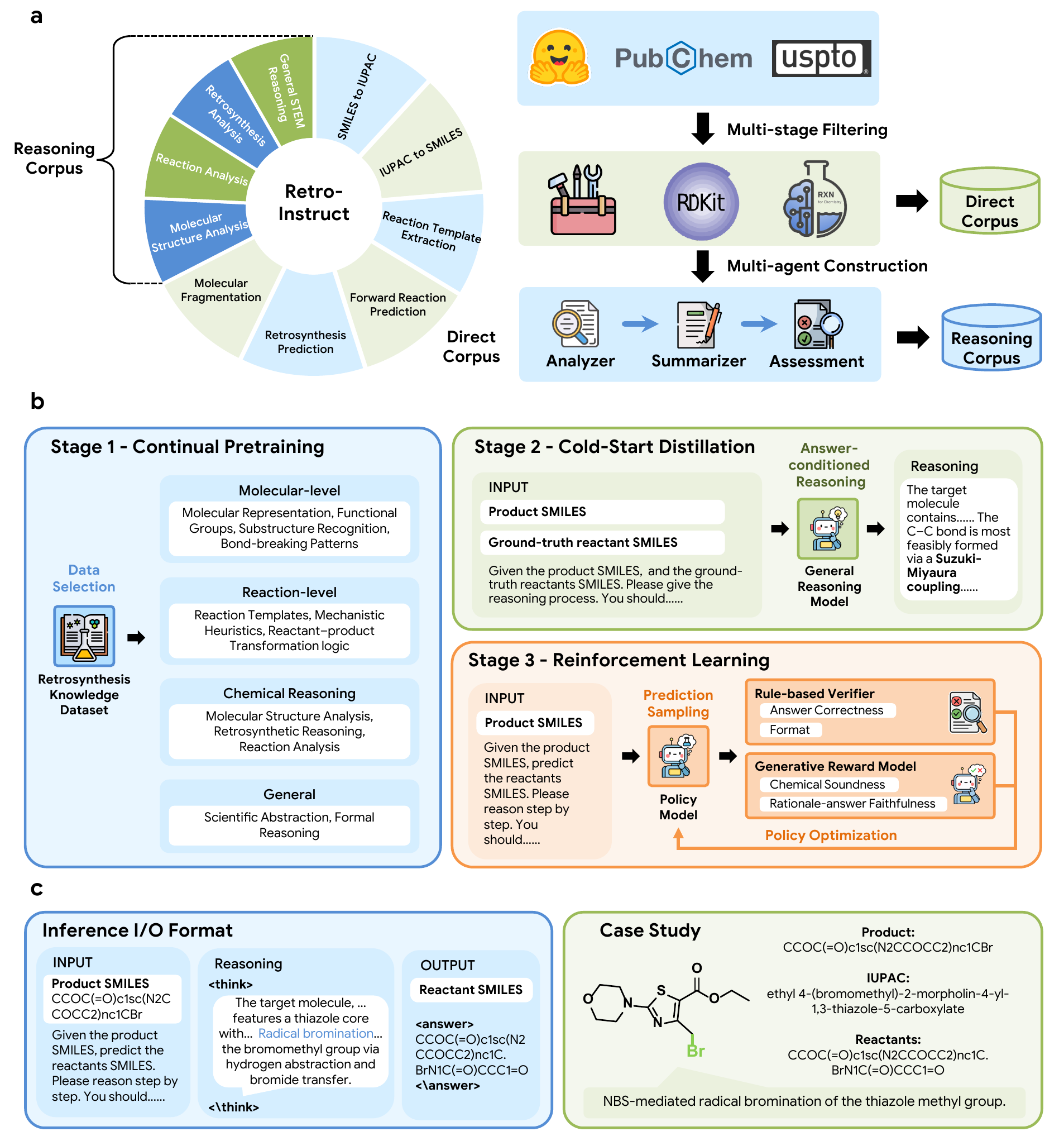}
    \caption{\textbf{Overview of \method{}.}
    \textbf{(a)} The composition and construction pipeline of the Retro-Instruct dataset.
    \textbf{(b)} The \method{} training framework comprises three stages: \textbf{Stage 1}, continual pretraining on Retro-Instruct; \textbf{Stage 2}, cold-start distillation to initialize and stabilize the model's reasoning capabilities; and \textbf{Stage 3}, large-scale reinforcement learning with both rule-based rewards and a generative reward model (GRM).
    \textbf{(c)} Illustration of \method{}'s reasoning for retrosynthesis, providing both explainable reasoning and the final predicted
reactants.
    }
    \label{fig:overview}
\end{figure}

To address these limitations, we introduce \method{}, a reasoning-driven LLM tailored for chemical retrosynthesis. \method{} integrates domain-specific chemical knowledge with explicit CoT reasoning to improve both prediction accuracy and model interpretability. By leveraging reinforcement learning and long-CoT training to elicit structured reasoning, \method{} effectively combines the generalization capabilities of LLMs with the precise requirements of synthesis planning.

Our approach has two main components. First, we curated \textbf{Retro-Instruct}, a high-quality dataset containing over 24 million retrosynthetic reaction samples across ten tasks. Through a rigorous multi-stage filtering pipeline, we removed noisy and chemically invalid records to ensure data consistency. This structured dataset enables \method{} to learn detailed molecular features and functional group transformations. Second, we developed a \textbf{three-stage training paradigm} to progressively enhance model capability: (1) \textit{Continual pretraining} on diverse tasks to establish a broad chemical knowledge base; (2) \textit{Cold-start distillation} to initialize CoT reasoning capabilities specific to retrosynthesis; and (3) \textit{Reinforcement learning}, using rule-based verifiable rewards for reactant accuracy and format compliance, together with a generative reward model (GRM) to improve the chemical soundness and faithfulness of the generated reasoning. The complete framework is outlined in \Cref{fig:overview}.

This combination of data-driven chemical knowledge and structured reasoning provides transparency in the prediction process, allowing chemists to inspect the rationale behind proposed retrosynthetic routes. We evaluate \method{} on the USPTO-50K and USPTO-FULL benchmarks, where it demonstrates superior performance over leading sequence-based and graph-based baselines. On USPTO-50K, \method{} achieves 60.4\% top-1 exact-match accuracy without test-time augmentation and 66.1\% under the full inference setup, with particular improvement in handling complex structures involving chirality and ring systems. Furthermore, double-blind human evaluations show a preference for our model's predictions over the baselines and confirm the high quality of its generated rationales. Finally, we demonstrate that \method{} can be applied to multistep retrosynthesis of real-world pharmaceuticals and functional organic materials, supporting its potential utility for practical synthesis planning. More broadly, \method{} provides a foundation for integrating explicit chemical reasoning into AI-assisted synthesis, with the potential to help chemists accelerate synthesis-route design and optimization.

\section{Results}
\subsection{Overview of \method{}}

The framework of \method{} is designed to emulate the deductive reasoning process of expert chemists (\Cref{fig:overview}). Unlike conventional end-to-end models, \method{} employs an LLM-based architecture optimized for synthetic reasoning. As depicted in the inference workflow, the model does not merely predict a reactant set; it first performs a granular structural analysis of the target molecule, identifying key functional groups and strategic disconnects before proposing chemically plausible precursors. This explicit reasoning mechanism serves as the core engine for improving both prediction accuracy and interpretability.

To support our reasoning-centric paradigm, we constructed \textbf{Retro-Instruct}, a comprehensive corpus of 24 million samples drawn from PubChem, USPTO, and general STEM reasoning data (\Cref{fig:overview}(a)). We first apply a multistage curation pipeline that uses cheminformatics tools to remove invalid molecules and noisy records. The resulting corpus comprises two complementary components. The direct-training corpus targets the core skills required for retrosynthesis: SMILES--IUPAC conversion transfers structural and functional-group information between human-readable chemical nomenclature and machine representations; reaction template extraction and molecular fragmentation emphasize reactive-pattern localization and disconnection priors; forward reaction prediction models reaction causality from reactants to products; and retrosynthesis prediction directly trains product-to-precursor mapping. In parallel, we construct three reasoning subsets covering molecular-structure analysis, reaction analysis, and retrosynthesis analysis, teaching the model to articulate chemical decisions as coherent rationales rather than only final answers. General STEM data further support broad question answering and instruction-following capabilities. Details are provided in \Cref{sec:retro-instruct}.

Building on this data foundation, we develop a three-stage training framework—\textbf{(1) continual pretraining, (2) cold-start reasoning distillation, and (3) large-scale reinforcement learning}, as illustrated in \Cref{fig:overview}(b). We first perform \textbf{continual pretraining} on Retro-Instruct to strengthen chemical representation understanding and acquire task-relevant priors. We then apply \textbf{cold-start reasoning distillation} to initialize structured retrosynthetic reasoning. Specifically, a general-domain teacher model is given the target product together with the ground-truth reactants and prompted to generate a step-by-step derivation (answer-conditioned distillation), which \method{} learns to imitate. Finally, we perform \textbf{reinforcement learning} using DAPO~\citep{yu2026dapo}. The composite reward combines rule-based rewards for reactant correctness and format compliance with generative reward model (GRM) assessments of the chemical soundness of the rationale and its consistency with the final reactants, thereby explicitly optimizing both prediction accuracy and reasoning quality.

During inference, \method{} accepts the SMILES representation of a target product and outputs a structured decision process. The rationale is enclosed within \texttt{<think>...\allowbreak</think>} tags, followed by the predicted reactants within \texttt{<answer>...\allowbreak</answer>} tags. The rationale identifies relevant functional groups, candidate reaction sites, and plausible transformations before the final reactants are generated (\Cref{fig:overview}(c)). This explicit structure facilitates human inspection of feasibility and precursor accessibility, and distinguishes \method{} from direct-output sequence models.

\subsection{Datasets}

We evaluate retrosynthesis performance on two widely used reaction datasets derived from U.S. patents (1976--2016): USPTO-50K~\citep{schneider2016s} and USPTO-FULL~\citep{dai2019gln}. USPTO-50K contains 50,016 manually curated organic reactions in SMILES format with reliable atom mapping between products and reactants, and is therefore commonly used as the primary benchmark for single-step retrosynthesis. In contrast, USPTO-FULL contains approximately 1 million reactions extracted directly from patent documents. While inherently noisier, its scale and diversity provide a complementary testbed for assessing model generalization.

For both datasets, we adopt the same train/validation/test splits as previous works~\citep{coley2017retrosim, dai2019gln, zhong2022rsmiles, zhong2023graph2edits, han2024editretro}. Owing to its rigorous curation, USPTO-50K serves as our primary benchmark for single-step retrosynthesis evaluation, while the larger and more diverse USPTO-FULL dataset provides a complementary benchmark for assessing scalability and generalization. Together, these datasets enable a comprehensive evaluation of model performance across different retrosynthesis settings. All results are reported under the reaction-class-unknown setting. To prevent test-set leakage, we removed from the training corpus all reactions whose products appear in the USPTO-50K~\citep{schneider2016s} or USPTO-FULL~\citep{dai2019gln} test sets. We further conducted a subset-level audit across all 10 components of our training corpus and identified no test data leakage. The complete audit procedure and results are provided in Supplementary Section F.4.

\subsection{\method{} enhances LLM performance on retrosynthesis task}

\begin{figure}[h]
    \centering
\vspace{-0.2cm}\hspace{-0.5cm}
    \includegraphics[width=\linewidth]{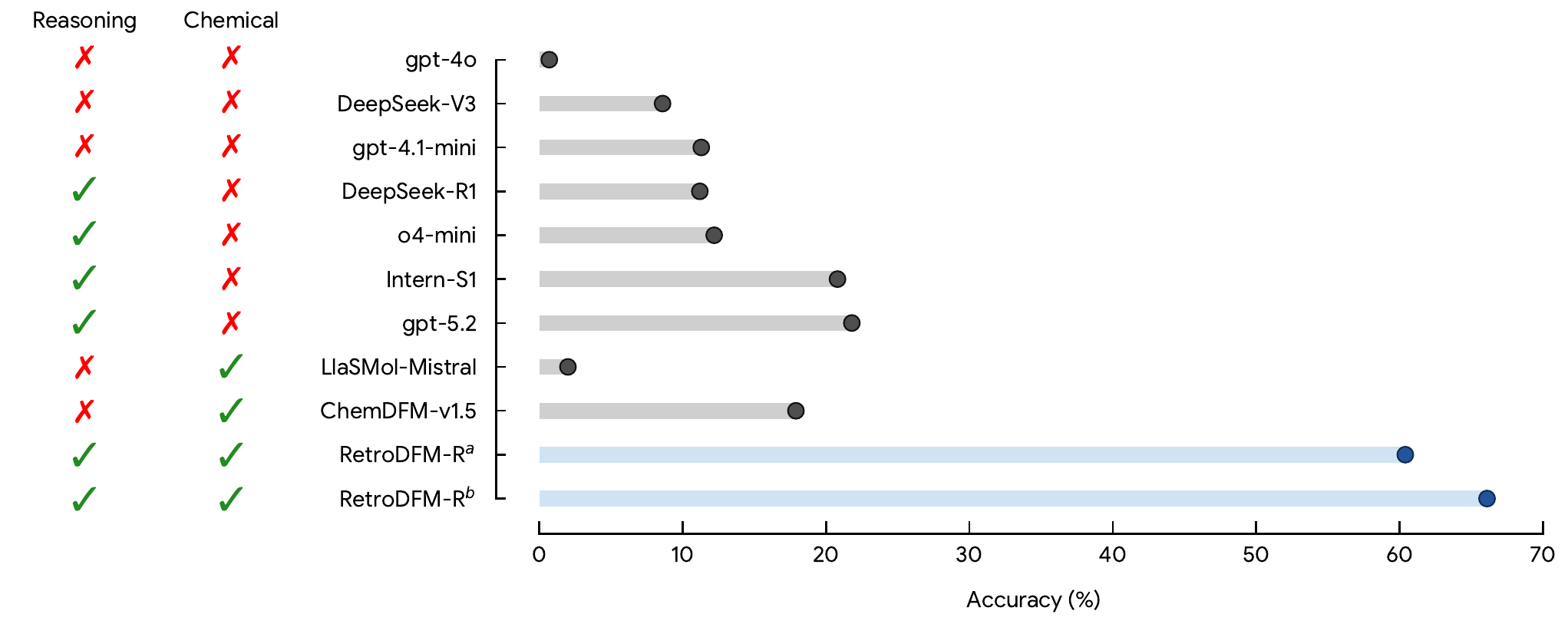}
    \caption{Performance comparison of different LLMs on the USPTO-50K benchmark.\\
\textsuperscript{a} Results for \method{} without test-time augmentation or repeated sampling.\\
\textsuperscript{b} \method{} top-1 accuracy with test-time augmentation, as discussed in \Cref{sec:main-results}.}
    \label{fig:llm-result}
\vspace{-0.2cm}
\end{figure}

We first benchmark \method{} against several state-of-the-art large language models, spanning general-domain non-reasoning, reasoning, and chemistry-specific LLMs.  Specifically, we evaluate: (1) \textbf{Non-reasoning LLMs}: gpt-4o~\citep{hurst2024gpt}, gpt-4.1-mini-2025-04-14~\citep{openai_gpt41}, and DeepSeek-V3-0324~\citep{liu2024deepseek}; (2) \textbf{Reasoning LLMs}: gpt-5.2-2025-12-11~\citep{gpt5_2}, o4-mini-2025-04-16~\citep{openai_o3_o4mini}, DeepSeek-R1-0526~\citep{guo2025deepseekr1} and Intern-S1~\citep{bai2025intern}; and (3) \textbf{Chemical LLMs}: ChemDFM-v1.5~\citep{zhao2025developing} and LlaSMol-Mistral~\citep{yu2024llasmol}. All models are evaluated on the USPTO-50K test set without test-time augmentation~\citep{tetko2020state, zhong2022rsmiles} or repeated sampling. Performance is measured using exact match accuracy, which compares the canonical SMILES of predicted reactants to the ground truth.

As shown in \Cref{fig:llm-result}, \method{} substantially outperforms all evaluated general-domain and chemistry-specific LLMs. Specifically, \method{} achieves an accuracy of 60.4\%, far exceeding the best-performing general-domain models. Among these models, reasoning-oriented LLMs generally outperform non-reasoning counterparts, with GPT-5.2 achieving the highest accuracy of 21.8\%, followed by Intern-S1 at 20.8\%. This performance gap suggests that general-purpose LLMs remain limited in their ability to handle specialized chemical representations and retrosynthetic reasoning. Among chemistry-specific models, ChemDFM-v1.5 reaches 17.9\%, indicating that chemical-domain pretraining improves domain familiarity but remains insufficient for accurate retrosynthesis without task-specific reaction supervision and explicit reasoning optimization. Overall, these results demonstrate that the strong performance of \method{} arises from combining chemical-domain adaptation with explicit retrosynthetic reasoning.

\subsection{Benchmarking \method{} against state-of-the-art methods}\label{sec:main-results}

\begin{table}[h]
\centering
\small
\setlength\tabcolsep{15pt}
\caption{Top-\textit{k} exact match accuracy (\%) on the USPTO-50K dataset under reaction-class unknown conditions.}
\label{tab:result-50k}
\begin{tabular}{lcccc}
\toprule
\multirow{2.5}{*}{\textbf{Model}} & \multicolumn{4}{c}{\textbf{Top-\textit{k} accuracy (\%)}} \\
\cmidrule{2-5}
 & \textbf{1} & \textbf{3} & \textbf{5} & \textbf{10} \\
\midrule
\rowcolor{lightblue!50} \multicolumn{5}{l}{\textit{Template-Based Methods}} \\
\quad RetroSim~\citep{coley2017retrosim} & 37.3 & 54.7 & 63.3 & 74.1 \\
\quad NeuralSym~\citep{segler2017neuralsym} & 44.4 & 65.3 & 72.4 & 78.9 \\
\quad GLN~\citep{dai2019gln} & 52.5 & 74.6 & 80.5 & 86.9 \\
\quad LocalRetro~\citep{chen2021localretro} & 53.4 & 77.5 & 85.9 & \textbf{92.4} \\
\midrule
\rowcolor{lightblue!50} \multicolumn{5}{l}{\textit{Graph-Based Methods}} \\
\quad G2Gs~\citep{shi2020graph} & 48.9 & 67.6 & 72.5 & 75.5 \\
\quad RetroXpert~\citep{yan2020retroxpert} & 50.4 & 61.1 & 62.3 & 63.4 \\
\quad MEGAN~\citep{sacha2021megan} & 48.1 & 70.7 & 78.4 & 86.1 \\
\quad GraphRetro~\citep{somnath2021learning} & 53.7 & 68.3 & 72.2 & 75.5 \\
\quad RetroPrime~\citep{wang2021retroprime} & 51.4 & 70.8 & 74.0 & 76.1 \\
\quad GTA~\citep{seo2021gta} & 51.1 & 67.6 & 74.8 & 81.6 \\
\quad G$^2$Retro~\citep{chen2023g} & 54.1 & 74.1 & 81.2 & 86.7 \\
\quad Graph2Edits~\citep{zhong2023graph2edits} & 55.1 & 77.3 & 83.4 & 89.4 \\
\midrule
\rowcolor{lightblue!50} \multicolumn{5}{l}{\textit{Sequence-Based Methods}} \\
\quad SCROP~\citep{zheng2019scrop} & 43.7 & 60.0 & 65.2 & 68.7 \\
\quad Aug.Transformer~\citep{tetko2020state} & 53.2 & -- & 80.5 & 85.2 \\
\quad Retroformer~\citep{wan2022retroformer} & 53.2 & 71.1 & 76.6 & 82.1 \\
\quad R-SMILES~\citep{zhong2022rsmiles} & 56.3 & 79.2 & 86.2 & 91.0 \\
\quad EditRetro~\citep{han2024editretro} & 60.8 & 80.6 & 86.0 & 90.3 \\
\quad \textbf{\method{}} & \textbf{66.1} & \textbf{84.2} & \textbf{88.6} & 92.2 \\
\bottomrule
\end{tabular}
\end{table}

\begin{table}[!h]
\centering
\small
\setlength\tabcolsep{15pt}
\caption{Top-$k$ exact match accuracy of \method{} and baseline models on the USPTO-FULL dataset under reaction-class unknown conditions.}
\label{tab:result-full}
\begin{tabular}{lcccc}
\toprule
\multirow{2.5}{*}{\textbf{Model}} & \multicolumn{4}{c}{\textbf{Top-\textit{k} accuracy (\%)}} \\
\cmidrule{2-5}
 & \textbf{1} & \textbf{3} & \textbf{5} & \textbf{10} \\
\midrule
\rowcolor{lightblue!50} \multicolumn{5}{l}{\textit{Template-Based Methods}} \\
\quad RetroSim~\citep{coley2017retrosim} & 32.8 & -- & -- & 56.1 \\
\quad NeuralSym~\citep{segler2017neuralsym} & 35.8 & -- & -- & 60.8 \\
\quad LocalRetro~\citep{chen2021localretro} &39.1 &53.3 &58.4 &63.7 \\
\quad GLN~\citep{dai2019gln} &39.3 & -- & -- &63.7 \\
\midrule
\rowcolor{lightblue!50} \multicolumn{5}{l}{\textit{Graph-Based Methods}} \\
\quad RetroXpert~\citep{yan2020retroxpert} & 49.4 & 63.6 & 67.6 & 71.6 \\
\quad MEGAN~\citep{sacha2021megan} &33.6 & -- & -- &63.9 \\
\quad Graph2Edits~\citep{zhong2023graph2edits} &44.0 &60.9 &66.8 &72.5 \\
\quad Graph2SMILES~\citep{tu2022permutation} &45.7 & -- & -- &63.4\\
\quad RetroPrime~\citep{wang2021retroprime} &44.1& -- & -- &68.5\\
\quad GTA~\citep{seo2021gta} &46.6 & -- & -- &70.4\\
\midrule
\rowcolor{lightblue!50} \multicolumn{5}{l}{\textit{Sequence-Based Methods}} \\
\quad Aug.Transformer~\citep{tetko2020state} & 46.2 & -- & -- &73.3\\
\quad R-SMILES~\citep{zhong2022rsmiles} &48.9 &66.6 &72.0 &\textbf{76.4}\\
\quad EditRetro~\citep{han2024editretro} &\textbf{52.2} &67.1 &71.6 &74.2\\
\quad \textbf{\method{}} & 51.3 & \textbf{67.6} & \textbf{72.2} & \textbf{76.4} \\
\bottomrule
\end{tabular}
\end{table}

To achieve optimal performance and a fair comparison with previous state-of-the-art retrosynthesis methods, we use the same SMILES augmentation method adopted by Retroformer~\citep{wan2022retroformer}, R-SMILES~\citep{zhong2022rsmiles}, and EditRetro~\citep{han2024editretro}. Because a molecule can be represented by multiple valid SMILES strings, we enumerate different SMILES roots to create alternative representations of the same product molecule. Second, we adopt a two-stage generation protocol to produce multiple candidate predictions. Specifically, we repeatedly sample intermediate reasoning trajectories and then the final reactant outputs, thereby maximizing the diversity of the proposed retrosynthetic routes. Further details of the inference procedure are provided in \Cref{sec:inference}.

We comprehensively evaluate \method{} on the USPTO-50K benchmark under the reaction-class-unknown setting, which reflects the practical scenario in which the reaction class is not provided. As shown in \Cref{tab:result-50k}, \method{} achieves a top-1 exact-match accuracy of 66.1\%, outperforming the previous best method, EditRetro, by 5.3 percentage points. The advantage extends to higher-rank predictions, with top-3, top-5, and top-10 accuracies of 84.2\%, 88.6\%, and 92.2\%, respectively. We further compare methods without test-time augmentation in the Supplementary Information. Even without beam search or other test-time augmentation, using only single-generation inference, \method{} achieves 60.4\% top-1 accuracy, surpassing the best sequence-based method, R-SMILES (53.2\%), by 7.2 percentage points and the best graph-based method, Graph2Edits (55.1\%), by 5.3 points. This result shows that \method{} produces strong first-choice predictions even in the simplest inference setting. Overall, these results demonstrate that a reasoning-centric LLM can outperform specialized retrosynthesis models while retaining an interpretable prediction process.

On the more challenging USPTO-FULL benchmark, the results of \method{} are summarized in \Cref{tab:result-full}. \method{} achieves a competitive top-1 accuracy of 51.3\% and establishes new state-of-the-art performance at top-3 and top-5, while tying the best reported top-10 accuracy.

The strong performance on USPTO-FULL highlights the ability of \method{} to handle a broad and diverse range of reaction types. Unlike template-based approaches, which are constrained by predefined reaction rules, \method{} can flexibly model retrosynthetic transformations without relying on a fixed template library. Compared with prior end-to-end template-free models, it further incorporates chemical knowledge and explicit reasoning to guide prediction, resulting in chemically plausible retrosynthetic solutions even on the diverse and noisy USPTO-FULL dataset. These results demonstrate that \method{} maintains strong generalization and chemical validity in complex, realistic retrosynthesis settings.

\subsection{Ablation study on multistage training}

To quantify the contribution of each stage in the training pipeline, we evaluate four sequential checkpoints on USPTO-50K: the initial Qwen3-8B model, continual pretraining, cold-start distillation, and reinforcement learning. Each checkpoint is evaluated under both single-generation inference and the full inference setup with test-time augmentation. This stage-wise comparison allows us to disentangle how chemical domain adaptation, reasoning distillation, and reinforcement learning contribute to the final retrosynthesis performance.

\begin{table}[h]
\centering
\small
\caption{Ablation study of the multistage training pipeline on USPTO-50K.}
\label{tab:stage-ablation}
\begin{tabular}{lccccc}
\toprule
\multirow{2}{*}{\textbf{Checkpoint}} & \textbf{Single} & \multicolumn{4}{c}{\textbf{Full inference setup}} \\
\cmidrule(lr){2-2}\cmidrule(lr){3-6}
 & \textbf{Acc.} & \textbf{Top-1} & \textbf{Top-3} & \textbf{Top-5} & \textbf{Top-10} \\
\midrule
Qwen3-8B & 0.0 & 0.0 & 0.0 & 0.0 & 0.0 \\
Continual pretraining & 30.4 & 55.0 & 77.4 & 84.9 & 90.0 \\
Cold-start distillation & 39.8 & 58.8 & 79.7 & 85.5 & 90.6 \\
Reinforcement learning & \textbf{60.4} & \textbf{66.1} & \textbf{84.2} & \textbf{88.6} & \textbf{92.2} \\
\bottomrule
\end{tabular}
\end{table}

\textbf{Continual pretraining.} The base Qwen3-8B model obtains 0.0\% accuracy, showing that general pretraining alone does not provide the specialized mapping from product SMILES to reactants. Continual pretraining raises the single-generation accuracy to 30.4\% and reaches 55.0\% top-1 and 90.0\% top-10 accuracy under the full inference setup. This stage therefore establishes the required chemical representations and reaction knowledge. The gap between top-1 and top-10 also indicates that the model can often generate the correct reactants among its candidates before it can reliably rank them first.

\textbf{Cold-start distillation.} Cold-start distillation further increases single-generation accuracy from 30.4\% to 39.8\% and full-setup top-1 accuracy from 55.0\% to 58.8\%. Beyond this quantitative improvement, its primary role is to introduce explicit retrosynthetic reasoning by teaching the model to connect molecular-structure analysis, reaction-site identification, and disconnection selection into a coherent rationale. The comparatively smaller gains at higher top-$k$ values are consistent with this role: cold-start distillation primarily establishes a structured reasoning process rather than directly optimizing the final prediction objective.

\textbf{Reinforcement learning.} Reinforcement learning produces the largest subsequent improvement, increasing single-generation accuracy from 39.8\% to 60.4\% and full-setup top-1 accuracy from 58.8\% to 66.1\%. The stronger gain at top-1 indicates that reinforcement learning substantially improves the model's ability to identify the leading retrosynthetic solution. Importantly, this stage optimizes not only reactant correctness and format compliance but also the quality of the generated rationale. The generative reward model (GRM) evaluates chemical soundness and rationale--answer consistency, and both reasoning-related reward signals improve during reinforcement learning (Supplementary Figure 11). Thus, reinforcement learning strengthens both predictive performance and the reliability of the accompanying reasoning.

Overall, the stage-wise ablation demonstrates complementary roles for the three training stages. Continual pretraining establishes chemical domain knowledge and retrosynthesis competence, cold-start distillation introduces explicit and structured reasoning, and reinforcement learning further improves both first-choice prediction quality and rationale reliability. Meanwhile, the gap between full-setup top-1 accuracy and single-generation accuracy progressively narrows across training, indicating that the model becomes increasingly capable of producing strong first-choice predictions without relying heavily on test-time augmentation. We provide additional ablation studies on direct-answer training to isolate the effect of rationale-based training from model scale, as well as on inference-time augmentation components, in Supplementary Section G.

\begin{table}[h]
\centering
\small
\setlength\tabcolsep{15pt}
\caption{Top-$k$ RoundTrip and MaxFrag accuracy of the \method{} and baselines on USPTO-50K dataset with reaction class unknown.}
\label{tab:round-trip}
\begin{tabular}{lcccc}
\toprule
\multirow{2.5}{*}{\textbf{Model}} & \multicolumn{4}{c}{\textbf{Top-\textit{k} accuracy (\%)}} \\
\cmidrule{2-5}
 & \textbf{1} & \textbf{3} & \textbf{5} & \textbf{10} \\
\midrule
 \rowcolor{lightblue!50}\multicolumn{5}{l}{\textit{RoundTrip}} \\
\quad LocalRetro~\citep{chen2021localretro} & \textbf{89.5} & \textbf{97.9} & \textbf{99.2} & -- \\
\quad MEGAN~\citep{sacha2021megan} & 82.0 & 89.9 & 91.7 & 94.0 \\
\quad GraphRetro~\citep{somnath2021learning} & 86.0 & 89.9 & 91.7 & 94.0 \\
\quad Graph2Edits~\citep{zhong2023graph2edits} & 85.9 & 93.5 & 95.1 & 96.4 \\
\quad EditRetro~\citep{han2024editretro} & 86.1 & 96.2 & 98.1 & 98.9 \\
\quad \textbf{\method{}} & 87.7 & 97.1 & 98.5 & \textbf{99.2} \\
 \midrule
\rowcolor{lightblue!50} \multicolumn{5}{l}{\textit{MaxFrag}} \\
\quad AugTransformer~\citep{tetko2020state} & 58.5 & 73.0 & 85.4 & 90.0 \\
\quad MEGAN~\citep{sacha2021megan} & 54.2 & 75.7 & 83.1 & 89.2 \\
\quad Graph2Edits~\citep{zhong2023graph2edits} & 59.2 & 80.1 & 86.1 & 91.3 \\
\quad R-SMILES~\citep{zhong2022rsmiles} & 60.5 & 82.7 & 88.8 & 92.5 \\
\quad EditRetro~\citep{han2024editretro} & 65.3 & 83.9 & 88.9 & 92.8 \\
\quad \textbf{\method{}} & \textbf{70.4} & \textbf{87.2} & \textbf{90.8} & \textbf{93.8} \\
\bottomrule
\end{tabular}
\end{table}

\subsection{\method{} achieves superior performance in predicting plausible reactants}
Because retrosynthesis is inherently a one-to-many task, a single product molecule may correspond to multiple possible reactant sets. To more faithfully assess prediction correctness under this setting, we evaluate the round-trip accuracy of \method{} using the pretrained forward-synthesis model, Molecular Transformer (MT)~\citep{schwaller2019molecular}, which tests whether the predicted reactants can be converted back into the original product. The top-$k$ round-trip accuracy results are summarized in \Cref{tab:round-trip}. \method{} achieves a top-1 round-trip accuracy of 87.7\%, significantly outperforming previous graph-based and sequence-based models. While slightly lower than the template-based method LocalRetro~\citep{chen2021localretro}, this difference is largely due to the more constrained generation space of template-based approaches. Substantial improvements are also observed for top-3, top-5, and top-10 accuracies compared to prior approaches. These results demonstrate that our CoT-enhanced retrosynthetic reasoning model more effectively captures chemical knowledge and generates plausible reactant predictions.

\subsection{\method{} accurately predicts the main reactants}
In addition to round-trip accuracy, we employ the MaxFrag accuracy metric to assess the exact match of the largest reactant in each prediction. This metric is specifically designed to mitigate the limitations introduced by ambiguous reagent annotations in the dataset. By focusing on the largest reactant, MaxFrag emphasizes the model's ability to predict the most critical precursor structures in synthetic pathways. The top-$k$ MaxFrag accuracy results are presented in \Cref{tab:round-trip}. As shown, \method{} achieves a substantial improvement in \mbox{top-1} accuracy over previous methods and consistently outperforms baselines across top-3, top-5, and top-10 metrics.

\subsection{\method{}'s reasoning process provides explainability and insights into chemical retrosynthesis}

\begin{figure}[!b]
    \centering
    \includegraphics[width=0.9\linewidth]{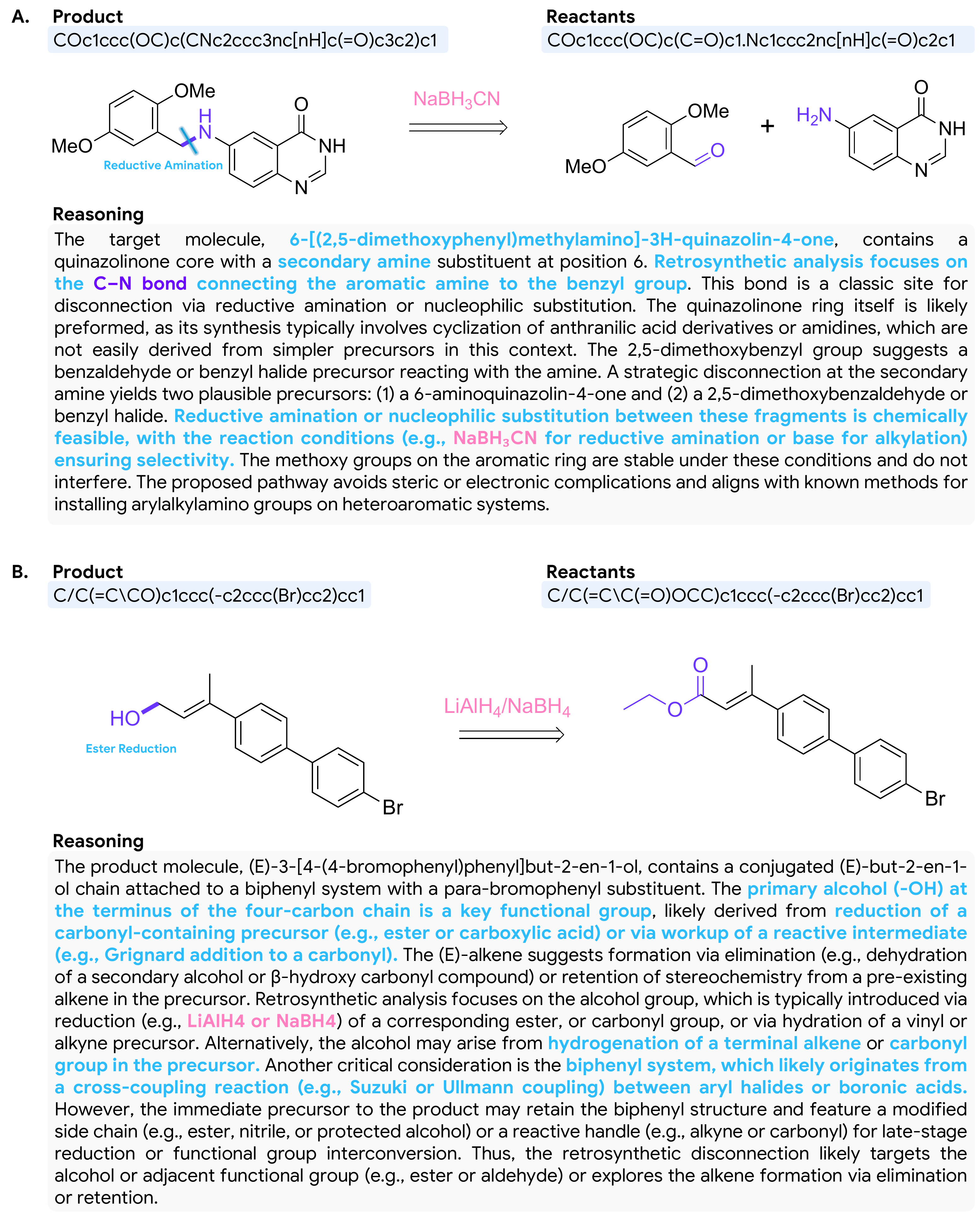}
    \caption{Step-by-step reasoning process generated by \method{}, highlighting key bond disconnections, reaction types, and the rationale for each retrosynthetic decision.}
    \label{fig:reason}
\end{figure}

A significant advantage of \method{} over previous retrosynthesis models is its ability to not only generate chemically feasible precursor reactants, but also provide clear and detailed rationales for its synthetic decisions. Leveraging the reasoning capabilities of large language models, \method{} provides human-readable explanations that allow chemists to inspect the model's proposed reaction center, disconnection, and transformation. This approach supports trustworthy AI-assisted retrosynthesis by making the prediction process more transparent and by offering practical chemical insights rather than only a final reactant string.

Similar to expert chemists, \method{} first analyzes the target molecule's structure and then identifies plausible retrosynthetic breakdown strategies, often referencing established reaction types such as named reactions or classical synthetic methods. Moreover, the model may propose multiple viable synthetic routes, along with potential catalysts and suitable reaction conditions required to realize these transformations.

We illustrate the reasoning capabilities and advantages of \method{} through two representative cases selected from the USPTO-50K dataset. \Cref{fig:reason}(A) illustrates an example where \method{} not only identifies the correct reaction type, but also explicitly recommends the appropriate reagent and reaction conditions. The model demonstrates clear and chemically sound reasoning in this case. It accurately assigns the IUPAC name of the target molecule and correctly pinpoints the key retrosynthetic disconnection at the methylamino bridge—the C–N bond linking the quinazolinone core to the aryl group. Recognizing this, \method{} proposes reductive amination as the optimal strategy to form the C–N bond. Importantly, the model further suggests sodium cyanoborohydride as a mild, chemoselective reducing agent commonly used in such transformations. This stepwise reasoning highlights both the explainability and practical chemical insight provided by \method{}, including explicit recommendations for key reagents and suitable reaction conditions—an aspect that conventional models often overlook. The ability to specify not only the reaction type but also the appropriate reagent (such as sodium cyanoborohydride) further demonstrates the practical superiority of reasoning-enabled language models in retrosynthetic planning.

\Cref{fig:reason}(B) showcases another example, highlighting \method{}'s ability to combine local disconnection analysis with longer-range synthetic planning. The model identifies the terminal primary alcohol as the most immediate retrosynthetic handle, considers plausible carbonyl-derived precursors, and suggests appropriate reducing reagents such as LiAlH$_4$ or NaBH$_4$. Beyond this local transformation, it further recognizes that the biphenyl scaffold could be assembled at an earlier stage through cross-coupling reactions, together with the corresponding catalytic chemistry. This example illustrates that \method{} can reason not only about the immediate disconnection and suitable reaction conditions, but also about upstream bond construction within a broader synthetic strategy.

\subsection{Performance on reactions with chirality, ring-forming and ring-opening}

\begin{figure}
    \centering
    \includegraphics[width=1\linewidth]{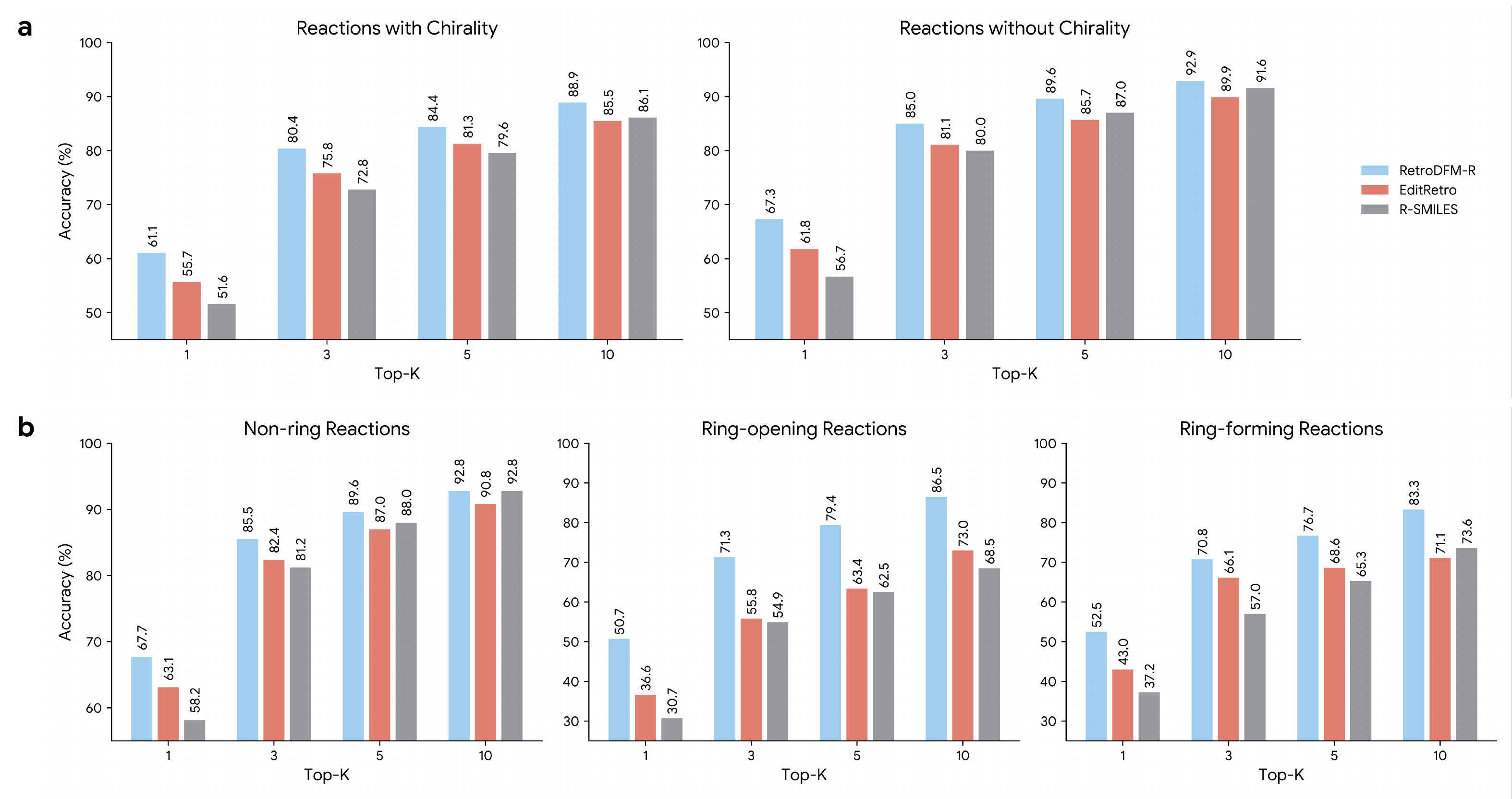}
    \caption{Top-$k$ performance of \method{} compared with baseline R-SMILES and EditRetro on complex reactions. (a) Reactions with or without chirality. (b) Reactions with non-ring, ring-opening and ring-forming. }
    \label{fig:comparison}
\end{figure}

Reactions involving stereochemistry, ring formation, and ring opening are particularly important in retrosynthesis because they often require precise control over molecular configuration and substantial changes in molecular topology. The stereochemical configuration of chiral centers can critically influence the biological activity and metabolic behavior of drug molecules, making stereochemistry-aware retrosynthesis an important challenge. Ring-forming and ring-opening transformations are likewise central to scaffold construction and functional-group interconversion, but often involve substantial bond rearrangements and topological changes that complicate retrosynthetic prediction.

\Cref{fig:comparison} summarizes performance on the USPTO-50K benchmark after stratifying reactions according to chirality and ring-related transformations. Across all categories, \method{} consistently outperforms EditRetro~\citep{han2024editretro} and R-SMILES~\citep{zhong2022rsmiles}. In particular, \method{} achieves higher top-1 accuracy for both chiral and non-chiral reactions, with the improvement on chiral cases highlighting its ability to reason about stereochemically constrained transformations. Similar gains are observed for ring-forming and ring-opening reactions, where successful prediction requires identifying appropriate bond disconnections and reconstructing substantial changes in molecular topology. These results indicate that the combination of chemical-domain pretraining and explicit reasoning improves the model's handling of structurally challenging transformations. The consistent gains on non-ring reactions further show that these improvements extend beyond specialized reaction classes and reflect broader advances in retrosynthetic prediction.

\subsection{Multistep retrosynthesis planning with \method{}}

\begin{figure}[!h]
    \centering
    \includegraphics[width=0.94\linewidth]{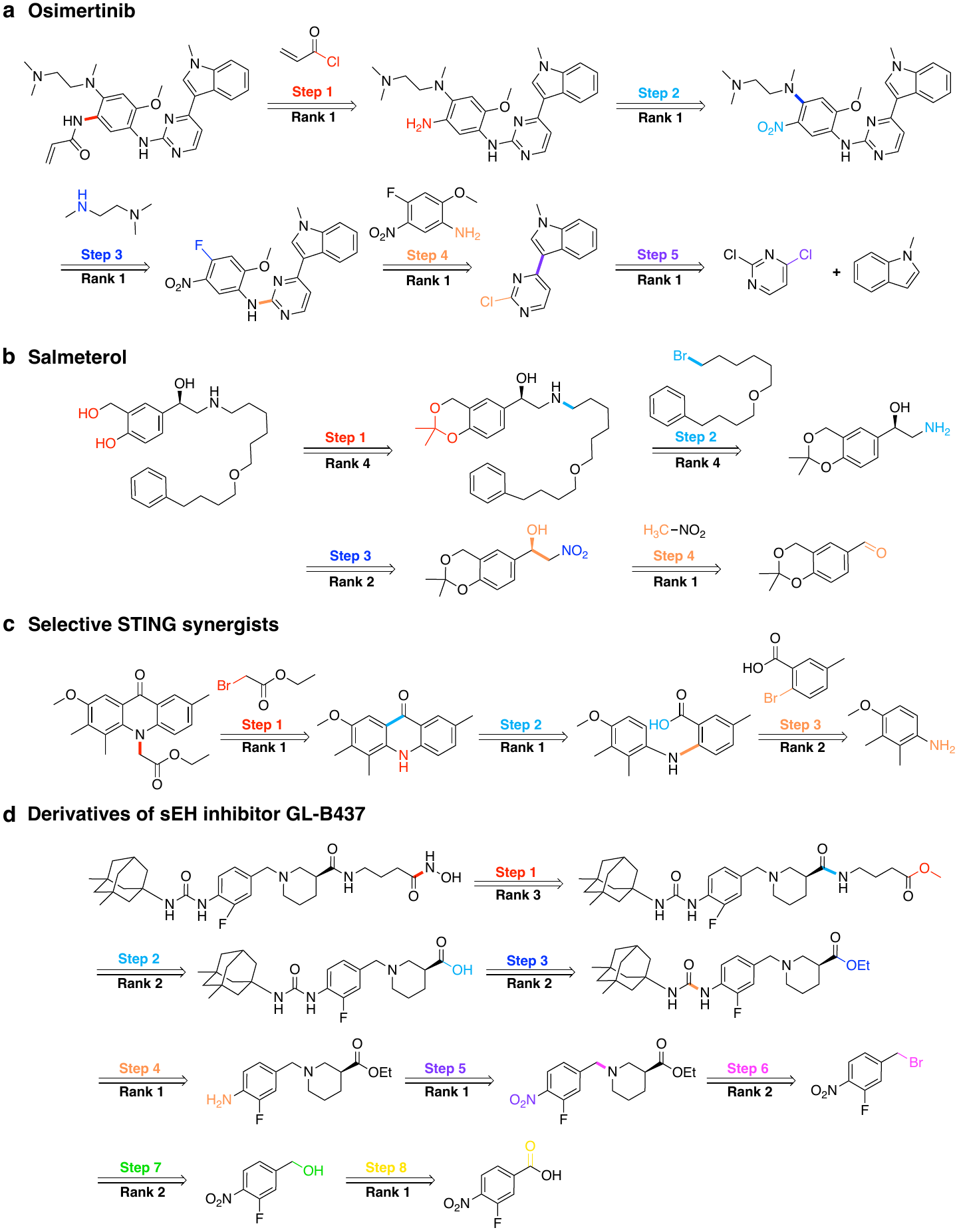}
    \caption{Multistep retrosynthesis of (a) Osimertinib, (b) Salmeterol, (c) a selective STING synergist, and (d) a derivative of the sEH inhibitor GL-B437, as predicted by \method{}. At each step, atoms and bonds involved in the transformation are highlighted.}
    \label{fig:multistep}
\end{figure}

To evaluate the practical utility of \method{} in real-world multistep retrosynthetic planning, we performed sequential retrosynthetic predictions using our single-step retrosynthesis model trained on the USPTO-50K dataset. Four pharmaceutically significant molecules were selected for detailed case studies: (1) Osimertinib~\citep{finlay2014discovery}, a third-generation EGFR tyrosine kinase inhibitor for treating EGFR-mutated non-small cell lung cancer; (2) Salmeterol~\citep{guo2011enantioselective}, a long-acting $\beta$2-adrenergic receptor agonist (LABA) for managing asthma and COPD; (3) a selective STING synergist with applications in cancer immunotherapy~\citep{hou2025design}; and (4) a dual-target derivative of the sEH inhibitor GL-B437 for treating inflammatory diseases~\citep{xu2024design}. The first two compounds have been previously evaluated in related studies~\citep{zhong2022rsmiles, zhong2023graph2edits, han2024editretro}, whereas the selective STING synergists and GL-B437 derivatives were selected from recent medicinal chemistry literature published in 2024 and 2025, respectively. Crucially, neither the final products nor their synthetic intermediates were present in the training data.

For \textbf{Osimertinib}, shown in \Cref{fig:multistep}(a), Finlay et al.~\citep{finlay2014discovery} described a five-step route from readily available starting materials. \method{} successfully predicted all five steps with rank-1 accuracy. Specifically, it identified the necessity of an initial acylation reaction with acryloyl chloride and accurately deduced subsequent nitro group reduction. \method{} further correctly predicted two sequential nucleophilic aromatic substitution (SNAr) reactions to introduce both an amino side chain and a nitroaniline moiety, matching the methodology of Finlay et al.~\citep{finlay2014discovery}. Finally, \method{} correctly proposed the Friedel--Crafts-type heteroarylation between N-methylindole and 2,4-dichloropyrimidine, accurately recovering the key C--C bond-forming disconnection in the reported synthesis.

For \textbf{Salmeterol}, whose synthetic pathway is illustrated in \Cref{fig:multistep}(b) and was initially evaluated by the R-SMILES study~\citep{zhong2022rsmiles}, \method{} successfully recovered all four literature steps at ranks four, four, two, and one. Importantly, the model accurately identified the final step as an asymmetric Henry reaction, a key transformation in the synthesis of Salmeterol~\citep{guo2011enantioselective}. Notably, our top-ranked prediction for this step precisely matched the reported experimental procedure, highlighting the model's ability to correctly handle reactions involving chirality.

\Cref{fig:multistep}(c) presents the synthesis of the \textbf{selective STING synergists} reported by Hou et al.~\citep{hou2025design} in 2025. \method{} correctly identified the synthetic route beginning with the reaction of an amine with ethyl bromoacetate, followed by the key ring-forming acylation, and culminating in an Ullmann reaction with 2-bromo-5-methylbenzoic acid. \method{} accurately predicted each step of this synthetic pathway, with all reactions correctly ranked within the top-2 predictions.

For the recently reported \textbf{GL-B437 derivative}, Xu et al.~\citep{xu2024design} described an eight-step synthetic route involving repeated transformations of carboxylic-acid derivatives. \method{} predicts all eight steps within the top three. Its top-1 predictions differ from the reported route only in minor choices of carboxylic-acid derivatives, such as using a free acid instead of a methyl ester or a methyl ester instead of an ethyl ester. These variations preserve the same underlying retrosynthetic logic. \method{} also recovers the final borane-mediated reduction of the carboxylic acid to the corresponding primary alcohol, demonstrating reliable multistep planning across closely related functional-group interconversions.

We further evaluate whether this capability extends beyond pharmaceutical targets by applying \method{} to self-assembled monolayer materials used in perovskite solar cells. As shown in Supplementary Section D, \method{} successfully predicts the synthetic routes of MPA-CPA and V1036. These results suggest that its route-planning capability is not limited to drug-like molecules, but also extends to functional organic materials.

\subsection{Assessing Retrosynthesis Quality via Double-Blind AB Test}
To rigorously evaluate the quality of retrosynthetic predictions from \method{}, we conducted double-blind AB tests against both the ground-truth labels in the dataset and the previous state-of-the-art method, EditRetro. For each comparison, we randomly sampled 100 reactions in which the predictions of \method{} differed from those of the respective reference. We invited three independent experts in organic chemistry to serve as annotators. For each reaction pair, two experts independently assessed the feasibility and chemical quality of the proposed retrosynthetic routes, choosing from four evaluation categories: “A is better”, “B is better”, “Both are equally good” or “Both are equally bad/unclear”. Detailed evaluation instructions are provided in Supplementary Section B.3. The AB tests were conducted in a double-blind manner, such that neither the annotators nor the experimenters knew the origin (model-generated or ground truth) of each retrosynthetic solution. In cases of disagreement between the first two evaluators, the third expert provided the final decision.

\begin{figure}[h]
    \centering
    \includegraphics[width=0.8\linewidth]{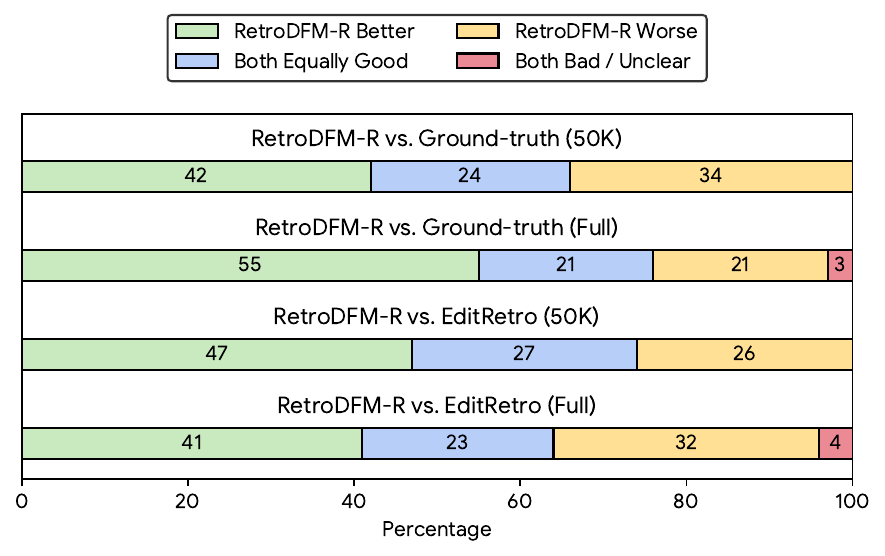}
    \caption{Human preference statistics for retrosynthesis predictions (AB tests).}
    \label{fig:preference}
\end{figure}

We present the AB-test preference results in~\Cref{fig:preference}. First, in comparison with the ground-truth reactants, 66\% of responses indicated that the outputs from \method{} were either preferred or considered equally good. Most of the observed discrepancies were due to minor substituent changes, such as chloride/bromine exchanges, which is reasonable given the one-to-many nature of retrosynthesis. This finding highlights \method{}'s strength in proposing chemically plausible alternatives, rather than merely replicating dataset answers. To further illustrate this, we provide 10 randomly selected examples in Supplementary Section G.4. In practice, the reactants predicted by \method{} are often indistinguishable from the ground truth in terms of reaction feasibility. On the noisier USPTO-FULL benchmark, where ground-truth labels may be less reliable due to atom-mapping errors or inaccurate reagent/reactant annotations, \method{}'s predictions are even more strongly favored (55\% preferred, 21\% equally good). These results highlight the model's robustness and its ability to propose chemically reasonable reactants even in the presence of noisy labels.

To further benchmark our approach, we compared \method{} with the previous state-of-the-art method, EditRetro\citep{han2024editretro}. On USPTO-50K, \method{} was consistently preferred over EditRetro in both preference and quality judgments. Similar trends were observed on USPTO-FULL, where both models' predictions were generally judged as comparable in quality. The superior performance of \method{} can be attributed to its ability to avoid trivial or invalid predictions, addressing a notable weakness of EditRetro. Collectively, these results demonstrate the effectiveness and reliability of \method{} in delivering high-quality, human-preferred retrosynthetic solutions.

We further evaluate the quality of inference-time rationales through a separate double-blind expert assessment. We randomly sampled 100 reactions from the USPTO-50K test set and collected rationales generated by \method{} and four strong reasoning LLMs: DeepSeek-R1, GPT-4.1-mini, o4-mini, and GPT-5.2. Each rationale was scored on a 1--5 scale across four dimensions: chemical soundness, clarity and conciseness, conclusion faithfulness, and expert-level insight. The complete rubric is provided in Supplementary Table 2. \method{} achieves the highest mean score across all four dimensions, including 4.21 for chemical soundness, 4.18 for conclusion faithfulness, and 3.69 for expert-level insight. Its chemical-soundness score slightly exceeds that of the strongest competing model, GPT-5.2, with larger gains in clarity, conclusion faithfulness, and expert-level insight. These results demonstrate that \method{} not only achieves strong predictive accuracy but also produces more reliable, interpretable, and expert-aligned rationales, supporting its use as a trustworthy tool for retrosynthetic analysis.

\section{Discussion}

We present \method{}, a reasoning-driven large language model designed for chemical retrosynthesis. \method{} bridges the gap between general-purpose, large-scale pretrained LLMs and the specialized requirements of retrosynthetic analysis. By adopting a step-by-step reasoning format, the model mirrors the workflow of expert chemists: it analyzes molecular structures, identifies plausible reaction sites and retrosynthetic disconnections, and can propose appropriate reagents, reaction conditions, or alternative synthetic routes. Importantly, \method{} not only generates chemically feasible precursor reactants, but also provides clear, human-readable rationales for its synthetic decisions.

We use a three-stage framework that progressively integrates domain adaptation with reasoning optimization. Continual pretraining on Retro-Instruct equips the model with domain-specific knowledge of molecular syntax, reaction transformations, and retrosynthetic mappings. Cold-start distillation then introduces structured retrosynthetic rationales, enabling coherent analysis of reaction sites and disconnection strategies. Reinforcement learning further refines this capability by jointly rewarding reactant correctness, format compliance, chemical soundness, and consistency between the rationale and the predicted reactants. The composite reward combines rule-based signals for answer correctness and formatting with generative reward model (GRM) evaluations of rationale quality and rationale--answer consistency. Together, these stages explicitly optimize both predictive accuracy and reasoning quality, allowing \method{} to generate accurate retrosynthetic predictions with interpretable and chemically grounded rationales.

Comprehensive experiments demonstrate the effectiveness of this design. On USPTO-50K, \method{} achieves 60.4\% top-1 accuracy without augmentation and 66.1\% with the full inference setup, outperforming previous state-of-the-art methods. Strong RoundTrip and MaxFrag scores further indicate that its predictions are often chemically plausible and preserve key precursor structures. In multistep planning, \method{} consistently recovers literature-reported disconnections for both medicinal-chemistry targets and functional organic materials. Beyond predictive accuracy, its explicit reasoning provides information on reaction types, reagents, conditions, and alternative routes, making the chemical basis of each prediction more transparent and easier to evaluate.

\begin{figure}[!t]
    \centering
    \includegraphics[width=0.88\linewidth]{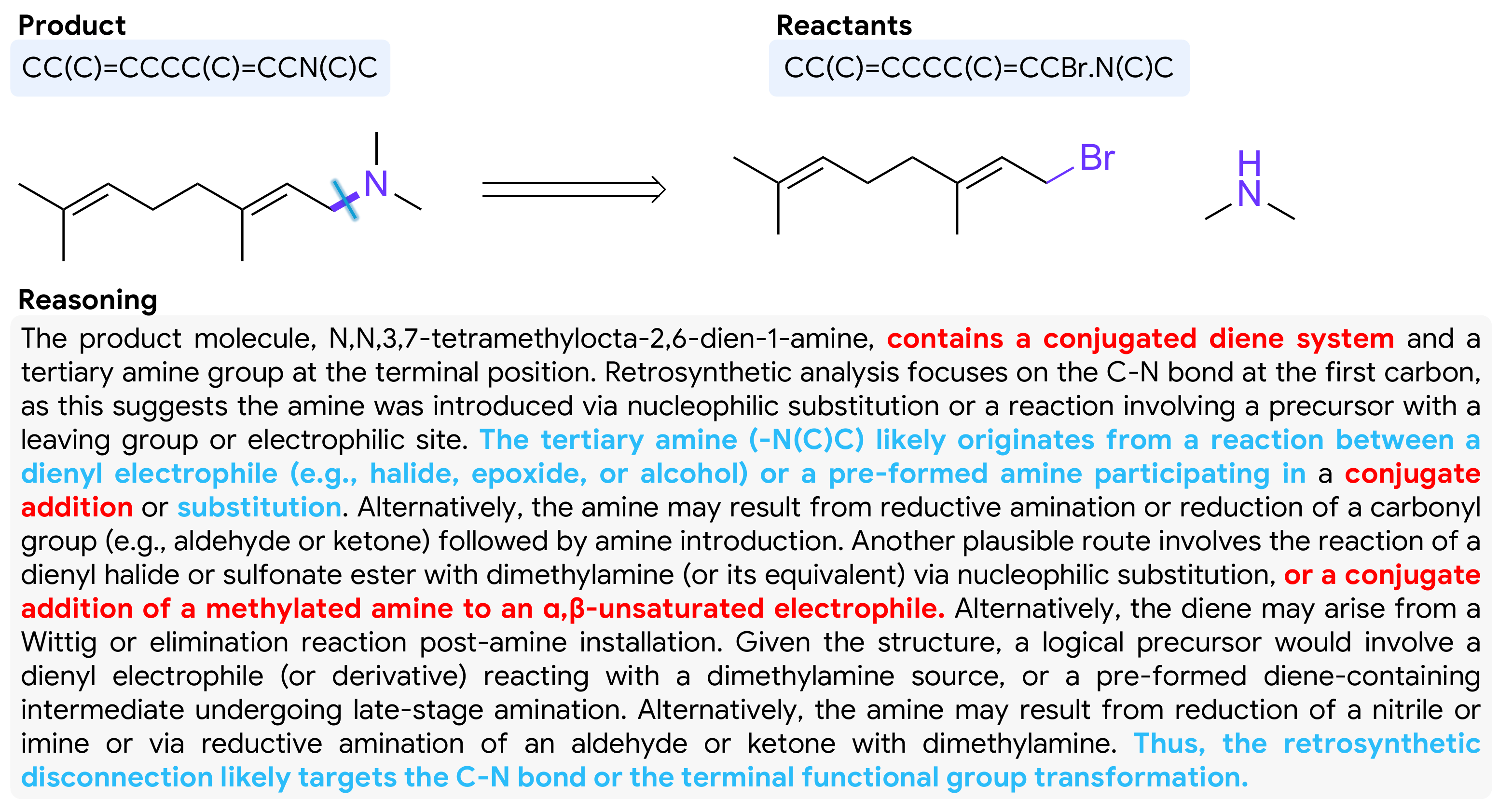}
    \caption{Representative structure-analysis error in a generated rationale. Erroneous rationale text is highlighted in red.}
    \label{fig:hallucination}
\end{figure}

Despite these promising results, several challenges remain. First, due to the probabilistic nature of large language models, \method{} may generate chemically incorrect or hallucinated rationales. For example, as shown in \Cref{fig:hallucination}, the model incorrectly identifies two non-conjugated double bonds as a conjugated system and consequently provides an inappropriate conjugate-addition explanation. Second, the current framework relies on one-dimensional SMILES representations for both input and output. Although SMILES is convenient and widely used for benchmark evaluation, it may not fully capture stereochemical, steric, and conformational effects that are more naturally represented in two or three dimensions. Incorporating multimodal molecular representations, together with high-quality reaction databases and chemical literature, could further ground the model's predictions and improve reliability for stereochemically sensitive transformations.

In summary, \method{} demonstrates the potential of reasoning-driven LLMs to jointly improve the accuracy and interpretability of retrosynthetic prediction. As foundation models and high-quality reaction databases continue to advance, we expect this line of work to accelerate the development of automated synthesis-planning systems that are both reliable and transparent.

\section{Methods}

This section details the complete framework of \method{}, encompassing data curation, model architecture, and the multi-stage training paradigm. We first describe the construction of \textbf{Retro-Instruct}, our large-scale instruction-tuning dataset. Subsequently, we outline the three-stage training pipeline, including continual pretraining, cold-start distillation, and reinforcement learning, and describe the inference setups. A schematic overview of the workflow is shown in \Cref{fig:overview}.

\subsection{Construction of Retro-Instruct}\label{sec:retro-instruct}

\textbf{Data Aggregation and Curation.}
To facilitate reasoning-driven retrosynthesis, we constructed \textbf{Retro-Instruct}, a composite dataset comprising 24,210,859 samples across ten tasks designed to cover diverse molecular and reaction-level skills. The dataset was constructed from three main sources: (1) reaction records from the USPTO datasets~\citep{schneider2016s,dai2019gln}, which provide the primary foundation for learning chemical transformations and retrosynthetic mappings; (2) molecular structures and corresponding IUPAC names from PubChem~\citep{kim2025pubchem}, which bridge human-readable chemical nomenclature and structural representations; and (3) 5,000 general STEM examples from NVIDIA's post-training dataset, consisting of exam-style multiple-choice questions to support instruction following and general scientific reasoning. Given the prevalence of noise in raw patent data, including incomplete reactant records and incorrect annotations, we applied a rigorous data-curation pipeline.

We first used RDKit to remove chemically invalid entries and molecules containing more than 200 atoms to ensure data quality and computational tractability. We then recomputed atom mappings for all reactions using RXNMapper and retained only reactions with a mapping confidence score above 0.3, ensuring reliable atom--atom correspondence. All molecular structures were canonicalized using RDKit before downstream processing.

To prevent test-data leakage, we further applied strict overlap filtering. Specifically, we extracted and canonicalized all product molecules from the USPTO-50K and USPTO-FULL test sets using RDKit 2025.3.2. We then removed any USPTO reaction whose product matched a test-set product, as well as any PubChem molecule overlapping with these test products. This filtering was performed before constructing both direct and reasoning task samples.

\textbf{Task Engineering.}
Building on the curated data, we constructed a suite of ten molecule- and reaction-level tasks, consisting of six direct tasks and four reasoning tasks. These tasks were designed to strengthen the model's understanding of molecular representations, reaction transformations, and retrosynthetic disconnections.

\begin{itemize}[leftmargin=12pt, itemsep=2pt, topsep=2pt]
\item \textbf{SMILES--IUPAC conversion:} We collected 10 million SMILES--IUPAC pairs from PubChem and constructed 10 million SMILES-to-IUPAC and 5 million IUPAC-to-SMILES samples. Given one representation, the model is asked to generate the corresponding representation in the other format. These tasks bridge human-readable chemical nomenclature and molecular syntax, while reinforcing the recognition of functional groups and molecular structures.

\item \textbf{Reaction template extraction:} We extracted atom-mapped reaction templates from USPTO reactions using RDChiral~\citep{Coley2019RDChiralAR}, yielding approximately 100,000 examples. Given a reaction SMILES, the model is asked to generate the corresponding atom-mapped reaction template. This task trains the model to identify reaction centers and abstract the local structural changes associated with chemical transformations.

\item \textbf{Molecular fragmentation:} We applied the BRICS algorithm~\citep{BRICS} implemented in RDKit to PubChem molecules, generating approximately 88,000 examples. Given a molecular SMILES, the model is asked to predict its BRICS fragments. This task exposes the model to synthetically meaningful bond-disconnection patterns.

\item \textbf{Retrosynthesis prediction:} This task aims to identify plausible reactant sets for a given target product. Using the cleaned and filtered USPTO reactions, we followed previous sequence-based retrosynthesis studies~\citep{tetko2020state,zhong2022rsmiles,han2024editretro} and applied \textbf{10-fold R-SMILES augmentation}, generating multiple non-canonical SMILES representations to improve robustness and generalization. Given a product SMILES, the model is trained to directly predict the corresponding reactants.

\item \textbf{Forward reaction prediction:} As the dual task of retrosynthesis, forward reaction prediction uses reactants as input and the corresponding product as output. We excluded single-reactant reactions, which provide limited context for learning reactant-to-product transformation patterns, resulting in approximately 460,000 examples. This task reinforces the model's understanding of reaction transformations and chemical causality.

\end{itemize}

\textbf{Synthetic Reasoning Generation.}
We additionally constructed three reasoning tasks---molecular-structure analysis, reaction analysis, and retrosynthesis analysis---to initialize the model's chemistry-related reasoning capabilities. The reasoning corpus was generated through three role-specific model calls. An \textit{Analyzer} first produced a detailed rationale following structured instructions and, where applicable, auxiliary chemical information in an answer-conditioned manner. A \textit{Summarizer} then refined the analysis into a concise and coherent rationale. Finally, an \textit{Assessment} agent evaluated each rationale for chemical correctness and logical consistency on a five-level scale, and only examples with an overall score of $\geq 4$ were retained.

For \textit{molecular-structure analysis}, we used molecules from PubChem and provided the Analyzer with both the SMILES and IUPAC name. The model was instructed to identify key functional groups and summarize the molecular connectivity, yielding 49,584 high-quality rationales after summarization and filtering. For \textit{reaction analysis}, we sampled reactions from the curated USPTO data and provided the reaction SMILES, asking the Analyzer to identify the key functional-group transformations and bond changes. This process yielded 85,316 reaction-analysis rationales. For \textit{retrosynthesis analysis}, we used the curated USPTO reactions and provided the product SMILES, IUPAC name, and ground-truth reactants. The Analyzer generated rationales covering product-structure analysis, reaction-site identification, and the proposed retrosynthetic transformation, resulting in 78,513 retained examples. We additionally constructed 122,812 retrosynthesis reasoning examples specifically for cold-start distillation.

For molecular-structure and reaction reasoning, both the Analyzer and Summarizer were instantiated with Qwen3-235B-A22B-Instruct, while MiniMax-M2.7 served as the Assessment agent. For retrosynthesis reasoning, we used DeepSeek-R1 as the Analyzer, DeepSeek-V3 as the Summarizer, and MiniMax-M2.7 for assessment. We further included 5,000 general STEM examples sampled from NVIDIA's Nemotron-SFT-Science-v2 dataset to support instruction following and general scientific reasoning. Together, the reasoning portion of Retro-Instruct contains 218,413 examples.

\textbf{Leakage Audit.}
To further verify the absence of test-set leakage, we conducted an exhaustive audit across all ten task categories in Retro-Instruct. We first extracted all product molecules from the USPTO-50K and USPTO-FULL test sets and canonicalized them using RDKit. For molecule-level tasks, including SMILES--IUPAC conversion and molecular fragmentation, every molecular SMILES appearing in each example was canonicalized and compared against the test-set products. For reaction-level tasks, including reaction-template extraction, retrosynthesis prediction, forward reaction prediction, and synthesis-related reasoning, we parsed each reaction, extracted and canonicalized its product, and checked for overlap with either USPTO test set. No overlapping examples were retained in the final training corpus. Further details are provided in Supplementary Section F.4.

Supplementary Section F provides the full data composition, reasoning-corpus construction procedure, model configurations, generation prompts, retained sample counts, and human evaluation of the synthesized reasoning subsets.

\subsection{Pipeline for \method{}}

As illustrated in \Cref{fig:overview}(a), the development of \method{} follows a progressive three-stage pipeline: (1) continual pretraining, (2) cold-start reasoning distillation, and (3) large-scale reinforcement learning.

\subsubsection{Stage 1: Continual Pretraining}
We continually pretrain Qwen3-8B on all 24 million Retro-Instruct examples with the standard autoregressive language-modeling objective. The direct tasks teach molecular syntax, product-reactant mappings, reaction-template abstractions, and forward reaction causality, while the reasoning tasks expose the model to structured chemical analysis. The SMILES--IUPAC conversion tasks also bridge human-readable chemical nomenclature and the SMILES strings used for model input and output, helping transfer structural and functional-group information into the representation needed for retrosynthesis. Continual pretraining therefore serves as domain adaptation before explicit rationale training, rather than only as additional language-modeling data. It runs for three epochs with a maximum sequence length of 16,384 tokens. The leakage audit described above is performed before training.

\subsubsection{Stage 2: Cold-Start Reasoning Distillation}
Cold-start distillation uses 122,812 answer-conditioned retrosynthesis examples. The teacher receives the product SMILES, IUPAC name, and ground-truth reactants and generates a rationale covering structure analysis, reaction-site identification, and the proposed transformation. This answer-conditioned setup is used because a general-domain reasoning teacher can produce coherent stepwise explanations but does not reliably solve difficult retrosynthesis questions from the product alone. Providing the correct reactants anchors the rationale to a chemically valid target while prompting the teacher to articulate the reasoning path as if deriving the precursors. The resulting demonstrations initialize the student with the desired \texttt{<think>} and \texttt{<answer>} structure before reinforcement learning. The student is trained for three epochs with standard supervised fine-tuning (SFT). The objective is the negative log-likelihood of the target sequence:
\begin{equation}
    \mathcal{L}_{\text{SFT}} = - \sum_{t=1}^{T} \log P_{\theta}(x_t \mid x_{<t}, \mathcal{I}),
\end{equation}
where $x$ represents the target sequence (reasoning chain and reactants), $\mathcal{I}$ denotes the input instruction and product, and $\theta$ represents the model parameters.

\subsubsection{Stage 3: Reinforcement Learning}
In the final stage, we applied large-scale reinforcement learning to further improve reasoning quality and retrosynthetic accuracy. To encourage exploration and increase input diversity, we augmented each product with multiple SMILES representations following previous protocols~\citep{zhong2022rsmiles,han2024editretro}, using 20-fold augmentation for USPTO-50K and 2-fold augmentation for USPTO-FULL. Molecular structures were first standardized with RDKit~\citep{Landrum2016RDKit} before generating the augmented SMILES. We then optimized the model using \textbf{DAPO} with a composite reward combining rule-based signals and generative reward model evaluations, as detailed in Section~\ref{sec:dapo}.

\subsection{Reinforcement Learning with DAPO}\label{sec:dapo}

We train \method{} with large-scale reinforcement learning using \textbf{DAPO}, combining rule-based, verifiable rewards with generative reward model (GRM) evaluations. The rule-based rewards directly assess reactant accuracy and output-format compliance, providing stable and objective training signals. In parallel, the GRM evaluates the generated rationale along two dimensions---chemical soundness and faithfulness to the predicted reactants---thereby explicitly optimizing both prediction correctness and reasoning quality.

Specifically, the rule-based component consists of $R_{\mathrm{answer}}$ and $R_{\mathrm{format}}$. For model output $o$ and ground-truth reactant set $a$, $R_{\mathrm{answer}}=1$ if the canonicalized predicted reactants exactly match $a$, and 0 otherwise. $R_{\mathrm{format}}=1$ if the output follows the required format
\lstinline[basicstyle=\ttfamily,breaklines=true]{<think>...</think>\n<answer>...</answer>},
and 0 otherwise.

To evaluate reasoning quality, we employ a GRM following recent work that formulates reward modeling through generation and explicit reasoning~\citep{zhang2025generative,liu2025inference,chen2026rmr,guo2025reward}. For each rollout, Qwen3.6-35B-A3B receives the product, complete rationale, and predicted reactants, and independently assigns two \textbf{binary} rewards. $R_{\mathrm{sound}}\in{0,1}$ evaluates whether the chemical statements, functional-group analysis, bond-disconnection reasoning, and proposed transformation are chemically correct and plausible. $R_{\mathrm{faith}}\in{0,1}$ evaluates whether the predicted reactants are consistent with and can realize the retrosynthetic strategy described in the rationale. Thus, faithfulness measures the observable agreement between the generated reasoning and the final prediction. Details of the GRM design and the complete evaluation prompt are provided in Supplementary Section E.1.

The final reward is defined as
\begin{equation}
R(o, a) = 1.0R_{\text{answer}}(o,a) + 0.2R_{\text{format}}(o)
+ 0.6R_{\text{faith}}(o) + 0.2R_{\text{sound}}(o),
\end{equation}
where $o$ contains the generated rationale and predicted reactants, and $a$ denotes the ground-truth reactant set. We assign the largest rationale-related weight to $R_{\mathrm{faith}}$, as consistency between the stated retrosynthetic strategy and the final reactants provides the most direct indication that the generated rationale genuinely supports the prediction.

We adopt the Dynamic sAmpling Policy Optimization (DAPO) algorithm~\citep{yu2026dapo} for reinforcement learning. DAPO is a policy-gradient method that improves upon the Group Relative Policy Optimization (GRPO) algorithm~\citep{shao2024grpo}. For each input question $q$, ground-truth answer $a$ from the data distribution $\mathcal{D}$, and a group of $G$ sampled outputs $\{o_i\}^G_{i=1}$, DAPO optimizes the following objective to improve the policy.
\begin{align}
    \mathcal{J}_{\text{DAPO}}(\theta) 
    &= \mathbb{E}_{(q,a)\sim \mathcal{D}, \{o_i\}_{i=1}^G\sim \pi_{\text{old}}(\cdot|q)} \left[
        \frac{1}{\sum_{i=1}^{G} |o_i|} \sum_{i=1}^{G} \sum_{t=1}^{|o_i|} 
        \min\left(
            r_{i,t}(\theta) \hat{A}_{i,t},\: c\cdot \hat{A}_{i,t}
        \right) 
    \right]  \\
    c &= \text{clip}\left(r_{i,t}(\theta), 1 - \epsilon_{\text{low}}, 1 + \epsilon_{\text{high}} \right).
\end{align}
where $\epsilon$ denotes the clipping range for the importance sampling ratio $r_{i,t}(\theta)$. The advantage $\hat{A}_{i,t}$ for the $i$-th response is computed by normalizing the group-level rewards ${R}_{i=1}^G$. The formulas for $r_{i,t}(\theta)$ and $\hat{A}_{i,t}$ are as follows:
\begin{equation}
 r_{i,t}(\theta) = \frac{\pi_\theta(o_{i,t} \mid q, o_{i, <t})}{\pi_{\text{old}}(o_{i,t} \mid q, o_{i, <t})}, \hat{A}_{i,t} = \frac{R_i - \text{mean}(\{R_i\}_{i=1}^G)}{\text{std}(\{R_i\}_{i=1}^G)}.
\end{equation}

DAPO extends Proximal Policy Optimization (PPO)~\citep{schulman2017proximal} and GRPO with asymmetric clipping. We use lower and upper clipping values of 0.20 and 0.28 and discard prompt groups whose rewards have zero variance. Supplementary Table 4 and Supplementary Section G.3 report the remaining RL configuration and the training dynamics for answer correctness, faithfulness, and soundness.

\subsection{Model architecture and training}
We build \method{} on Qwen3-8B~\cite{yang2025qwen3}, a dense decoder-only Transformer~\citep{vaswani2017attention} with 36 layers, model width $d_{\text{model}}=4096$, feed-forward dimension $d_{\text{ffn}}=12,288$, and a vocabulary of 151,936 tokens. It uses grouped-query attention~\citep{ainslie-etal-2023-gqa} with 32 query heads and 8 key/value heads. All stages use full-parameter training on one node with eight NVIDIA H200 GPUs. Continual pretraining uses the Megatron backend in SWIFT with tensor parallelism and sequence parallelism. Cold-start SFT uses SWIFT with DeepSpeed ZeRO-2. RL uses slime with Megatron policy training, SGLang rollout generation, and distributed GRM evaluation. Continual pretraining takes approximately 40 hours, cold-start distillation approximately 5 hours, and RL approximately 7 days for USPTO-50K and 10 days for USPTO-FULL. The configurations needed to reproduce each stage are provided in Supplementary Section E; lower-level launcher options are available in the released code.

\subsection{Inference augmentation}\label{sec:inference}
To generate diverse predicted reactants for each input SMILES, previous end-to-end sequence-based approaches typically apply beam search directly over reactant SMILES~\citep{tetko2020state,zhong2022rsmiles,han2024editretro}. In contrast, as an LLM-based model, \method{} generates a longer structured response before producing the final reactants. Direct beam search over the full output space can therefore produce candidates that differ mainly in the intermediate text while converging to identical reactant predictions, limiting diversity at the answer level.

To address this issue, we adopt a two-stage inference augmentation strategy that explicitly promotes diversity at the answer generation stage with repeated sampling. In the first stage, the model generates multiple intermediate reasoning trajectories for a given product molecule. In the second stage, conditioned on each reasoning trajectory, the model samples final reactant SMILES from the \texttt{<answer>} tag using a higher sampling temperature, which encourages exploration of alternative but chemically plausible reactant sets.

Specifically, we generate $k_s$ distinct reasoning trajectories per input. For each trajectory, we then sample $k_b$ reactant predictions by increasing the sampling temperature during answer generation (empirically set to 1.4 in experiments). This procedure concentrates diversity on the final reactant outputs rather than on variations in the reasoning text, resulting in a total of $k_s \times k_b$ reactant predictions per input molecule.

Following the inference-time augmentation strategy used in Augmented Transformer~\citep{tetko2020state} and R-SMILES~\citep{zhong2022rsmiles}, we further apply randomized SMILES representations. Specifically, for each product in the test set, we generate $k_a$ distinct SMILES representations by varying the root atom and traversal order. Combined with the two-stage generation procedure, this yields $k_a \times k_s \times k_b$ reactant predictions for each product. All generated reactant SMILES are canonicalized. Invalid SMILES strings are removed, and duplicate predictions are merged. For each unique canonical reactant set $y \in \mathcal{Y}$, we define its frequency as
\begin{equation}
\text{Count}(y) = \sum_{t=1}^{k_a k_s k_b} \mathbb{I}\left[\hat{y}_t = y\right],
\end{equation}
where $\{\hat{y}_t\}_{t=1}^{k_a k_s k_b}$ denotes the set of canonicalized predictions obtained from augmented inference, and $\mathbb{I}[\cdot]$ is the indicator function. We then rank the unique predictions in $\mathcal{Y}$ by $\text{Count}(y)$ in descending order and select the top-$k$ candidates for evaluation. The full augmented inference setting uses $(k_a,k_s,k_b)=(20,10,10)$. Ablations of each inference component and their effects on top-$k$ accuracy are provided in Supplementary Section G.2.

\section{Data availability}
All data used in this study are obtained from publicly accessible sources, including PubChem (\url{https://pubchem.ncbi.nlm.nih.gov}), the USPTO reaction dataset (\url{https://github.com/Hanjun-Dai/GLN}), and Nemotron-SFT-Science-v2 from NVIDIA's Nemotron post-training collection (\url{https://huggingface.co/datasets/nvidia/Nemotron-SFT-Science-v2}). Nemotron-SFT-Science-v2 is distributed under the CC BY-SA 4.0 license. The evaluation dataset is available at \url{https://huggingface.co/datasets/OpenDFM/retrodfm-R-inference}.

\section{Code availability}
The source code and the \method{} checkpoint are available at \url{https://github.com/OpenDFM/RetroDFM-R}.

\bibliography{ref.bib}
\bibliographystyle{naturemag}

\section{Acknowledgements}
We thank Z. Yu for valuable suggestions regarding the evaluation of multistep retrosynthesis. We gratefully acknowledge computing resources on the Zhiyuan-1 cluster provided by the Network \& Information Center at Shanghai Jiao Tong University, which supported part of the computational experiments in this work. This work was supported by the National Science and Technology Major Project (2023ZD0120703), the China NSFC Projects (62120106006, 92370206, 62576212, and U23B2057), and the Frontier Technologies R\&D Program of Jiangsu (BF2025029, and BF2025061).

\section{Author contributions}
S.Z., H.L., and L.C. conceived the research. S.Z. and H.L. conducted the experiments and drafted the manuscript. S.Z. was responsible for data processing. X.L. analyzed the results. Z.H.Z., Z.C.Z. and B.C. provided suggestions on experimental design and manuscript revision. D.L. contributed to the preparation of the dataset. L.C., X.C., and K.Y. supervised the research. All authors reviewed and approved the final manuscript.

\section{Competing interests}
The authors declare no competing interests.

\end{document}